\documentclass[conference]{IEEEtran}
\IEEEoverridecommandlockouts
\usepackage{cite}
\usepackage{amsmath,amssymb,amsfonts}
\usepackage{algorithmic}
\usepackage{graphicx}
\usepackage{textcomp}
\usepackage{xcolor}
\usepackage{hyperref} 
\usepackage{array}
\usepackage{subfigure}
\usepackage{amsfonts}
\usepackage{hyperref}
\usepackage{amsmath}
\usepackage{tabularx}
\usepackage{bbding}
\usepackage{doi}
\usepackage{placeins}

\def\BibTeX{{\rm B\kern-.05em{\sc i\kern-.025em b}\kern-.08em
    T\kern-.1667em\lower.7ex\hbox{E}\kern-.125emX}}
\begin{document}

\title{Improving Acoustic Scene Classification in Low-Resource Conditions\\
\thanks{\IEEEauthorrefmark{1}Corresponding author}
\thanks{This work was supported by the National Natural Science Foundation of China under Grant No. 62276153, and by a grant from the Tsinghua University-Tsingshang Joint Institute for Smart Scene Innovation Design.}
}

\author{\IEEEauthorblockN{Zhi Chen\IEEEauthorrefmark{2}, Yun-Fei Shao\IEEEauthorrefmark{3}, Yong Ma\IEEEauthorrefmark{2}\IEEEauthorrefmark{1}, Mingsheng Wei\IEEEauthorrefmark{2}, Le Zhang\IEEEauthorrefmark{2}, Wei-Qiang Zhang\IEEEauthorrefmark{3}\IEEEauthorrefmark{1}}
\IEEEauthorblockA{\IEEEauthorrefmark{2}Jiangsu Normal University, Xuzhou 221116, China \\
\IEEEauthorrefmark{3}Department of Electronic Engineering, Tsinghua University, Beijing 100084, China \\
\tt 2020221340@jsnu.edu.cn, wqzhang@tsinghua.edu.cn}}

\maketitle

\begin{abstract}
Acoustic Scene Classification (ASC) identifies an environment based on an audio signal. This paper explores ASC in low-resource conditions and proposes a novel model, DS-FlexiNet, which combines depthwise separable convolutions from MobileNetV2 with ResNet-inspired residual connections for a balance of efficiency and accuracy. To address hardware limitations and device heterogeneity, DS-FlexiNet employs Quantization Aware Training (QAT) for model compression and data augmentation methods like Auto Device Impulse Response (ADIR) and Freq-MixStyle (FMS) to improve cross-device generalization. Knowledge Distillation (KD) from twelve teacher models further enhances performance on unseen devices. The architecture includes a custom Residual Normalization layer to handle domain differences across devices, and depthwise separable convolutions reduce computational overhead without sacrificing feature representation. Experimental results show that DS-FlexiNet excels in both adaptability and performance under resource-constrained conditions.
\end{abstract}
 

\section{Introduction}

Acoustic Scene Classification (ASC) categorizes environments such as parks, airports, and streets based on their soundscapes. Deep learning, particularly Convolutional Neural Networks (CNNs) \cite{Morocutti2022RECEPTIVEFR,Zheng,Koutini2019}, have significantly enhanced classification accuracy. However, CNN models typically demand considerable hardware resources, posing challenges for deployment on resource-constrained devices. Additionally, performance inconsistency across unfamiliar devices arises due to variations in microphone quality and characteristics.

Early approaches used acoustic features such as Zero Crossing Rate, Perceptual Linear Prediction, and Mel Frequency Cepstral Coefficients (MFCCs) for classification\cite{Rakotomamonjy,Valero}. More recently, handcrafted features, including MFCCs and time-frequency representations, have been employed as inputs for Deep Neural Networks (DNNs) and CNNs \cite{ghodasara2016acoustic,mcdonnell2020acoustic}.

In practical applications, sounds are often recorded using different devices, making inter-device variability a significant challenge. To address this, data augmentation techniques such as SpecAugment, Mixup, Device Impulse Response, and audio mixing have been employed \cite{takahashi2016deep,hu2020device}.

To tackle these challenges, we propose DS-FlexiNet, which integrates depthwise separable convolutions from MobileNetV2 \cite{Sandler2018MobileNetV2IR} with ResNet-like residual connections. Quantization Aware Training (QAT) reduces model size and computational requirements without compromising accuracy. Additionally, data augmentation techniques like Auto Device Impulse Response (ADIR) and Freq-MixStyle (FMS) \cite{Kim2022DomainGW,Zhou2021DomainGW} further enhance cross-device generalization. Knowledge Distillation (KD) from an ensemble of teacher models also improves performance on previously unseen devices.

% Our research focuses on adapting models to low-resource conditions, as highlighted by DCASE 2022 Task 1A, which demands generalization to new devices with constraints on model size (128 KB) and MACs (30M). This promotes the practical deployment of complex models in real-world scenarios. 

\section{Methods}

This section introduces a new design for ASC, focused on cross-device performance in low-resource conditions.

\subsection{Network Architecture}

\begin{figure}
    \centering
    \includegraphics[width=0.5\textwidth]{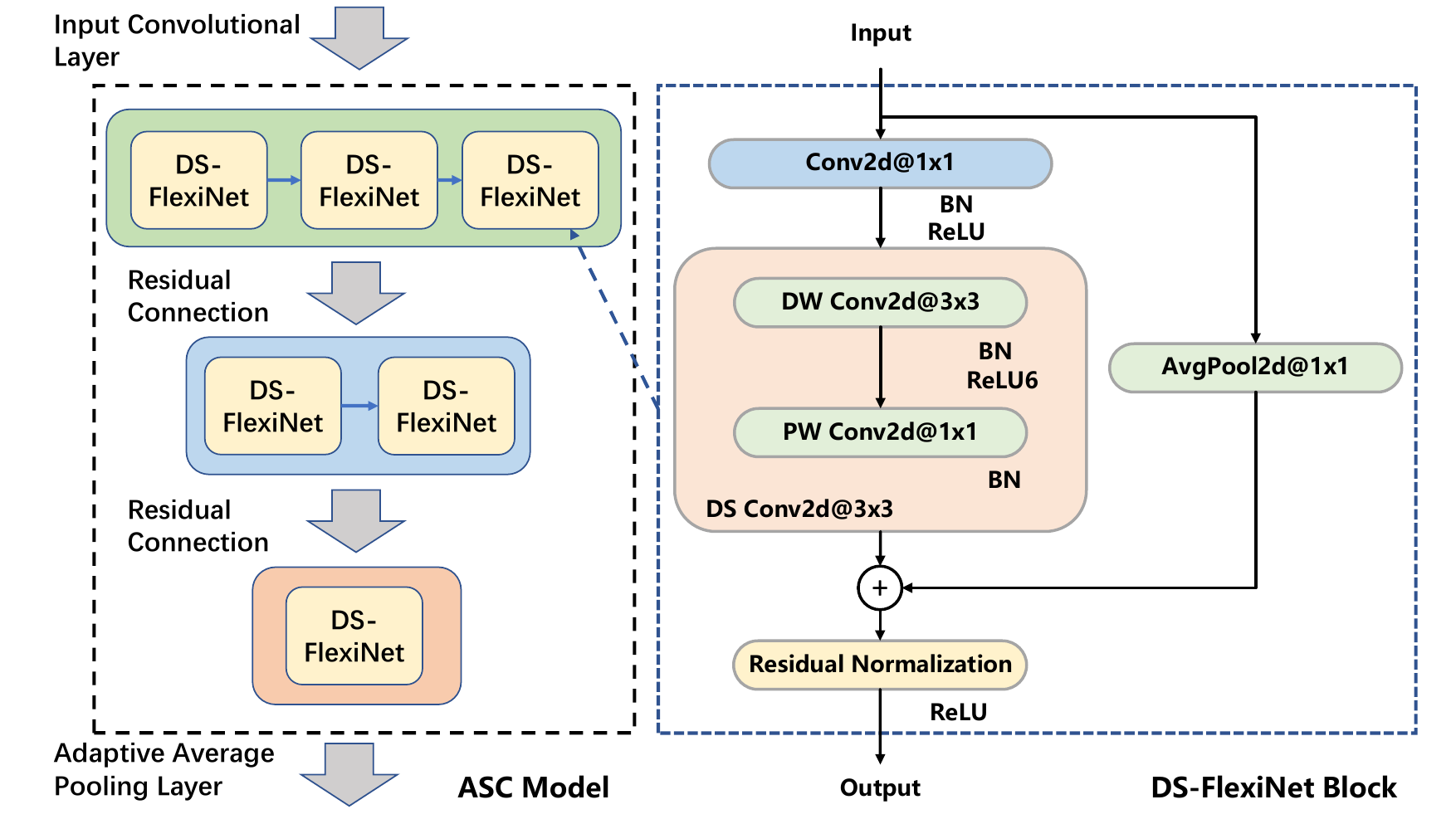}
    \caption{Network Architecture} 
    \label{fig:Network Architecture}
\end{figure}

The overall network architecture, inspired by MobileNetV2 \cite{Sandler2018MobileNetV2IR} and ResNet \cite{He2015DeepResidualLearning} principles, consists of multiple stages featuring DS-FlexiNet blocks. It starts with two 3x3 convolutions with stride 2, followed by Batch Normalization (BN) and ReLU activation, which reduces feature map size and complexity while enhancing abstraction. As stages advance, channel count increases, boosting representational capacity. Residual connections are used to preserve and add original inputs to post-convolution outputs, adapting based on channel and stride differences to improve information flow, training efficiency, and convergence speed. The model uses depthwise separable convolutions, which divide standard convolution into depthwise and pointwise operations \cite{Sandler2018MobileNetV2IR}. This approach lowers computational complexity and parameter count while preserving feature expressiveness and reducing memory usage. The complete architecture is shown in Fig. \ref{fig:Network Architecture}.

\subsection{Residual Normalization}

Instance Normalization (IN) is commonly used in image processing to minimize domain differences for improved domain generalization or style transfer \cite{Ulyanov2016TextureNF}. It works by normalizing channel means and variances.

 In ASC, we apply Instance Normalization to the channel dimension to capture audio device characteristics \cite{Shao2024DeepSL}. Unlike frequency-wise normalization, which operates across the frequency dimension \cite{Kim2022QTI}, our approach focuses on the channel dimension. This method helps reduce domain differences across various audio devices by normalizing the characteristics specific to each channel.

\begin{equation}
\mathrm{IN}(x)=\frac{x-\mu_{nc}}{\sqrt{\sigma_{nc}^2+\epsilon}}\label{equation:eq1}
\end{equation}

\begin{flushleft}

where, $\mu_{nc}$ and $\sigma_{nc}^2\in\mathbb{R}^{N\times C}$ represent the mean and variance of the input feature $x\in\mathbb{R}^{N\times C\times F\times T}$, computed along the channel dimension. The dimensions $N$, $F$, and $T$ correspond to the batch size, frequency, and time, respectively. $\epsilon$ is a small constant added to $\sigma_{nc}^2$ to avoid division by zero.

\end{flushleft}

Residual normalization (RN) preserves domain-specific information in acoustic features by using adaptive residual learning, which reduces domain discrepancies \cite{mcdonnell2020acoustic}.

\begin{equation}\mathrm{ResNorm}(x)=\lambda\cdot x+\mathrm{IN}(x)\end{equation}

\begin{flushleft}
where, $\lambda$ is a trainable parameter, allowing normalization to adjust to data characteristics and task requirements. 
\end{flushleft}

\subsection{Data Augmentation}

Data augmentation enhances model generalization, mitigates overfitting, and compensates for limited training data~\cite{takahashi2016deep}. 

\subsubsection{Freq-MixStyle}

FMS is a variant of MixStyle that operates on the frequency dimension rather than the channel dimension \cite{Kim2022DomainGW,Zhou2021DomainGW}. It normalizes the frequency bands in the spectrogram and utilizes the mixed frequency statistics for reverse normalization. This approach allows the model to incorporate frequency information when learning mixed features between samples, enhancing the model's robustness and performance in audio data processing. The application probability and mixing coefficients of FMS can be adjusted using hyperparameters.

\subsubsection{Auto Device Impulse Response}

ADIR improves model performance by adapting impulse response simulations to maintain high accuracy on unseen devices. By convolving audio signals with Device Impulse Responses (DIRs), the model becomes more robust to reverberation and generalizes better to unseen devices  \cite{Morocutti2023DeviceRobustAS}. 
The method involves randomly selecting one of 66 DIRs from $\mathrm{MiclRP}$, convolving it with the waveform. MicIRP devices have distinct frequency responses, making them suitable for simulating various recording environments.

\begin{figure}
    \centering
    \includegraphics[scale=0.36]{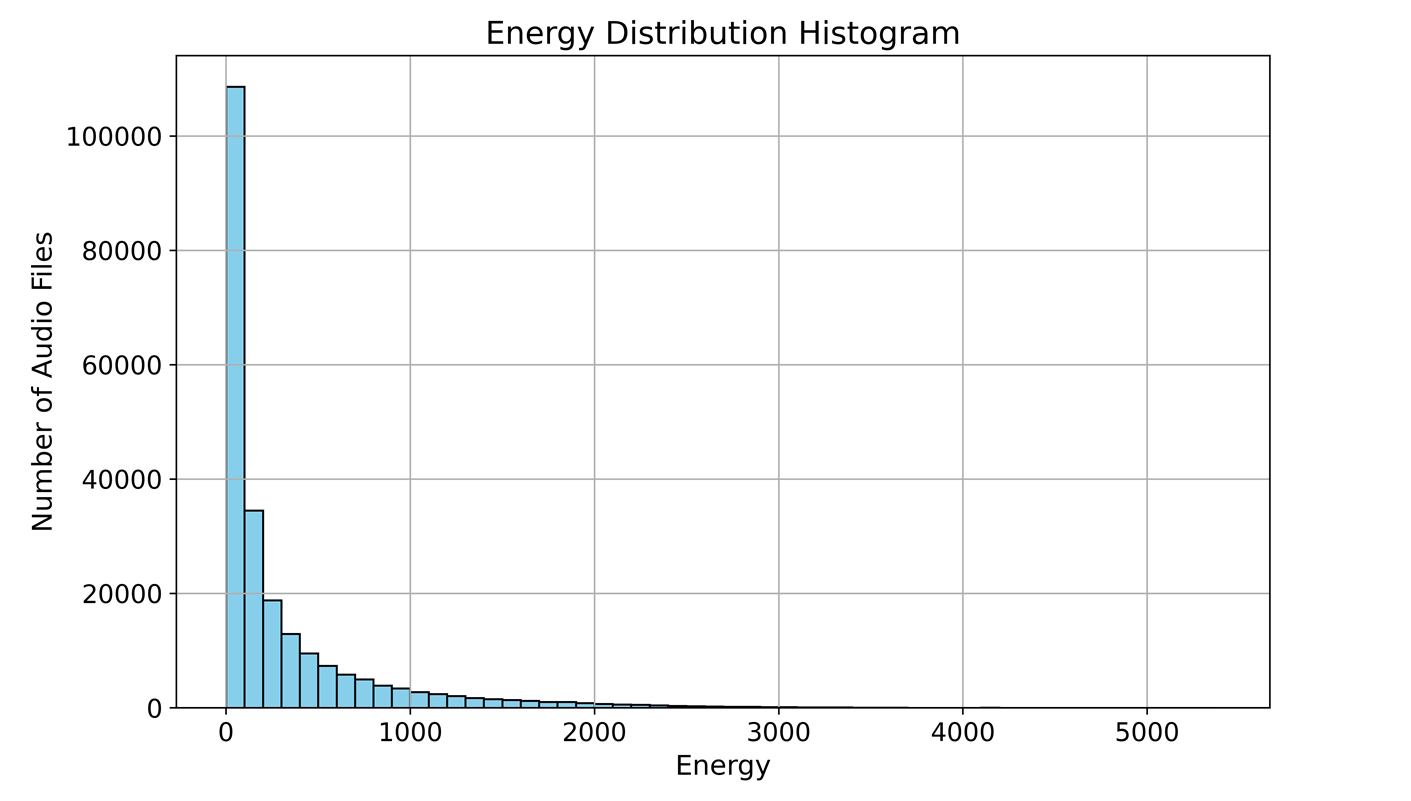}
    \caption{Energy Distribution Analysis of Audio in the TAU22 Dataset.} 
    \label{fig:Energy_distribution}
\end{figure}

Building on this concept, we propose ADIR method. By evaluating the energy level of audio samples, we adjust the convolution duration based on the audio's energy level. For high-energy audio samples, we apply impulse response techniques to preserve intricate characteristics while avoiding distortion. For low-energy samples, we skip impulse response processing to prevent overprocessing and potential distortion.  We analyzed energy distribution in the TAU22 dataset (see Fig. \ref{fig:Energy_distribution}). ADIR includes a hyperparameter $p_{A}$ to control the probability of DIR convolution.

\begin{equation}\begin{aligned}&\mathrm{ADIR}(x)=\begin{cases}x,&\text{if }\mathrm{E}(x)\le \mathrm{E}_{\mathrm{threshold}}\\\mathrm{IR}(x),&\text{otherwise}\end{cases}\\
&\mathrm{IR}(x)=x \otimes h_{\mathrm{dir}}
\end{aligned}\end{equation}

\begin{flushleft}

where, $x$ represents the input audio waveform, $\mathrm{E}(x)$ denotes its energy level, and $\mathrm{E}_{\mathrm{threshold}}$ is a predefined threshold. Based on energy distribution analysis, we set $\mathrm{E}$ to an average value of 323. $\mathrm{ADIR}(x)$ denotes the enhanced audio, while $h_{\mathrm{dir}}$ represents a randomly selected device impulse response (DIR) from the dataset $\mathrm{MiclRP}$.

\end{flushleft}

\subsection{Teacher Models Fusion }

In our study, we used KD to improve the student model. For the teacher models, we used a set of twelve PaSST and CP-ResNet models\cite{SchmidCPJKUST}. Building on this approach, we proposed an enhancement where 12 teacher models were constructed, each independently trained on the same dataset. The logits output from each model reflects its classification confidence for different acoustic scenes. To leverage the information from all teacher models, we developed a fusion strategy to integrate these logits into a soft target. The fusion formula is as follows: 

\begin{equation}h(i,t)=\sum\limits_{k=1}^{12}\alpha_k\cdot\text{logits}_{k}(i,t)+\beta_i\end{equation}
\begin{flushleft}

where, $h(i,t)$ is the output after merging the teacher model, with ${i}$ denoting the ASC task category and ${t}$ representing different audio inputs. The weight parameter ${\alpha_k}$ varies across system models, and $\mathrm{logits}_{(k)}(i,t)$ refers to the logits for category ${i}$ and input ${t}$ in system ${k}$. The bias parameter $\beta_i$ differs by category.

\end{flushleft}

% The total loss can be represented as:

Knowledge distillation effectively compresses models and enhances generalization in ASC tasks. The loss function comprises two components: label loss, $\mathrm{Loss}_{\mathrm{label}}$, which measures the discrepancy between predictions and ground truth, and distillation loss, $\mathrm{Loss}_{\mathrm{kd}}$, which captures the difference between teacher and student outputs. Both components use cross-entropy. The total loss is represented as:

\begin{equation}\mathrm{Loss}=\lambda \mathrm{Loss}_{\mathrm{label}}+(1-\lambda)\mathrm{Loss}_{\mathrm{kd}}\end{equation}
 
\begin{flushleft}
where, $\lambda$ is a weight-balancing hard label loss and distillation loss, updated during training. 

\end{flushleft}

\section{Experiment setup}

\subsection{Dataset}

We use Task 1A data from the TAU Urban Acoustic Scene 2022 Mobile Development dataset \cite{Heittola2020AcousticSC}, comprising 64 hours of recordings from 12 European cities across 10 acoustic scenes and 4 device types: 3 real (A, B, C) and 6 simulated (S1-S6). Additionally, the challenge organizer provides metadata indicating a training set of 139,620 samples and a test set of 29,680 samples. 

As shown in Table \ref{tab:TAU22}, devices S4, S5, and S6 are only present in the test set, meaning their specific characteristics are unseen during training.
\begin{table}[h]
\centering
\caption{TAU Urban Acoustic Scene 2022 Mobile Development dataset}
\label{tab:TAU22}
\begin{tabular}{>{\raggedright\arraybackslash}p{0.12\linewidth}>{\raggedright\arraybackslash}p{0.17\linewidth}>{\raggedright\arraybackslash}p{0.17\linewidth}>{\raggedright\arraybackslash}p{0.15\linewidth}>{\raggedright\arraybackslash}p{0.15\linewidth}}
\hline
\textbf{DEVICE} & \textbf{Total} & \textbf{Train} & \textbf{Test} & \textbf{Not used }\\
\hline
A& 144000& 102150& 3300& 38550\\
B,C& 10780+10770& 7490+7480& 2×3290&   —\\
S1,S2,S3& 3×10800& 3×7500& 3×3300&    —\\
S4,S5,S6& 3×10800&    —& 3×3300& 22500\\
\hline
Total& 230350& 139620& 29680&    —\\
\hline

\end{tabular}

\end{table}

\subsection{Experimental Preparation}

% For DS-FlexiNet, we use a 32 kHz sampling rate to compute a 256-bin Mel spectrogram with a 3072-length Hann window and an FFT hop size of 500. We compute the power spectrum, map it to Mel frequency space with 256 filters, and apply a logarithm to get the log Mel spectrogram. A 1-second audio file yields 64 frames of log Mel spectrogram with 256 frequency bins. Data augmentation includes random time-domain rolling and frequency masking.

% The teacher model uses PaSST and CP-ResNet fusion. Models are trained on GPUs with a batch size of 256 and cross-entropy loss. Training is done with PyTorch Lightning, with adjustable parameters for DS-FlexiNet. Audio data is preprocessed into log Mel spectrograms with data augmentation and KD to improve generalization. Networks are trained for 250 epochs. 

 For DS-FlexiNet, we use a 32 kHz sampling rate to compute a 256-bin Mel spectrogram. Each 1-second audio file produces 64 frames. Data augmentation includes random time-domain rolling and frequency masking.

The teacher model combines PaSST and CP-ResNet. It is trained with a batch size of 256 and cross-entropy loss using PyTorch Lightning. DS-FlexiNet training involves adjustable parameters and lasts for 250 epochs.

\subsection{Quantization Aware Training}

In this study, we optimized our deep learning model using PyTorch's quantization techniques, specifically QAT with the 'fbgemm' observer, to convert all parameters and computations in the low-complexity DS-FlexiNet model to int8.

\begin{figure*}[h]
\centering  %居中
\subfigure[RN on the channel dimension]{   %第一张子图
\begin{minipage}{0.52\linewidth}
\centering    %子图居中
\includegraphics[scale=0.45]{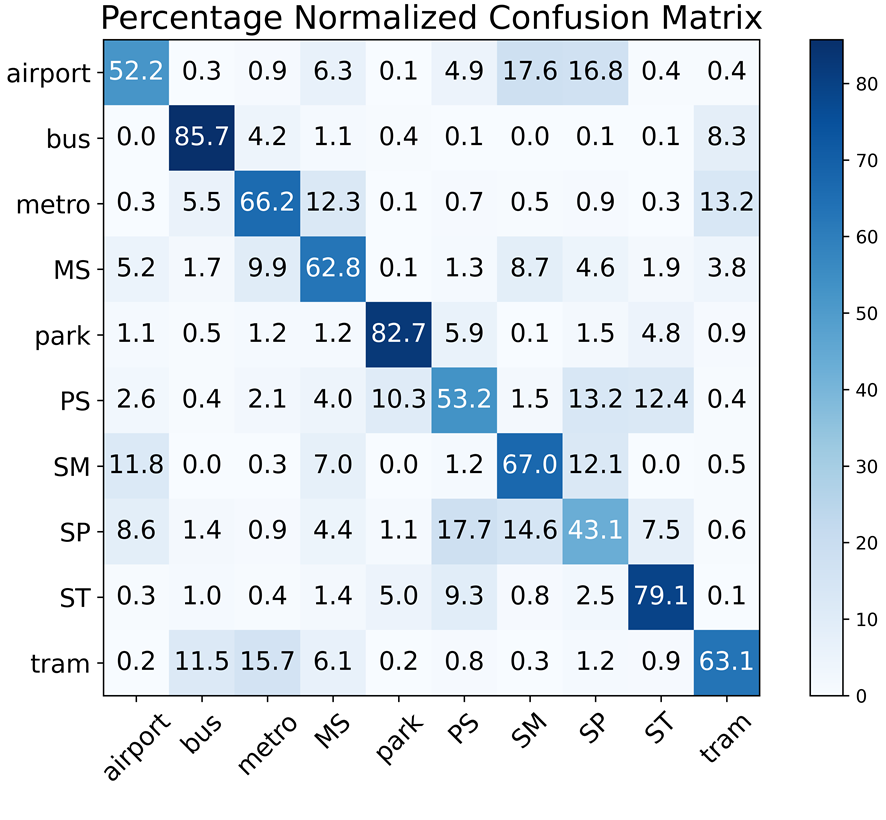}  %以jpg的0.5倍大小输出
\end{minipage}}
\subfigure[RN on the frequency dimension]{ %第二张子图
\begin{minipage}{0.45\linewidth}
\centering    %子图居中
\includegraphics[scale=0.45]{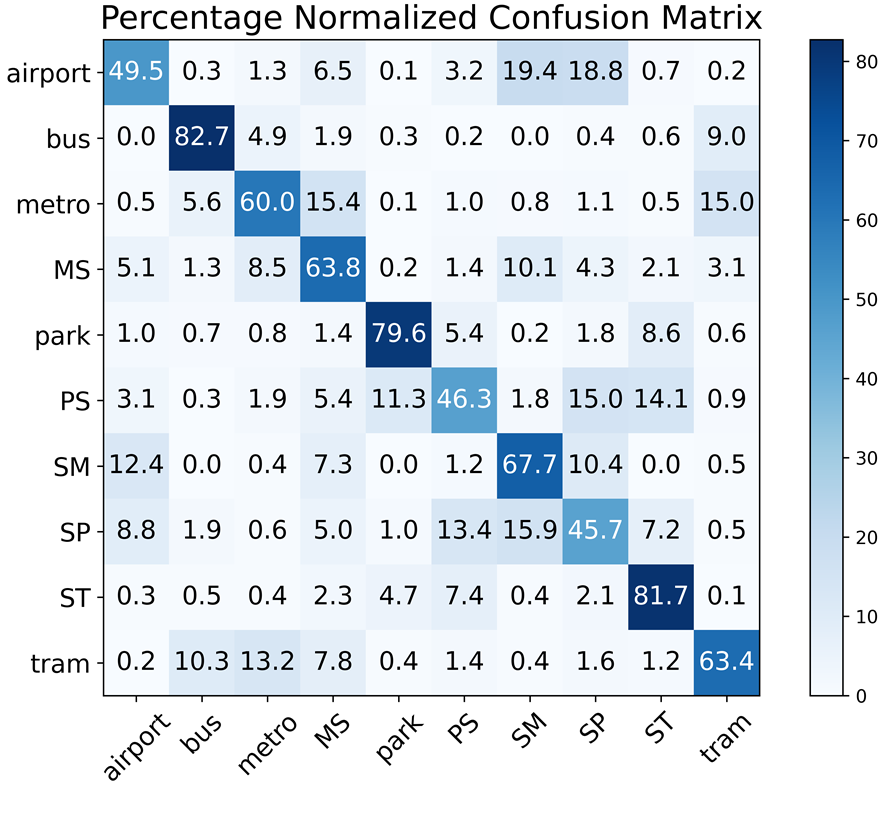}%以jpg的0.5倍大小输出
\end{minipage}}
\caption{The confusion matrix classified by different dimensions: MS:metro\_station, PS: public\_square, SM: shopping\_mall, SP: street\_pedestrian, ST: street\_traffic. The comparison experiment is based on the sm4 model.}    %大图名称
\label{fig:RN}    %图片引用标记
\end{figure*}

\section{Results and Analysis }

% This chapter primarily analyzes the model's outstanding capability in addressing acoustic scenes under resource-constrained environments and cross-device conditions. 

\subsection{Proposed Models }

This study proposes four student models, sm1 to sm4, to tackle resource constraints and cross-device conditions in acoustic scene recognition. Table \ref{tab:proposed-models} details their parameter configurations and performance. To enhance performance in low-resource conditions, we applied QAT to convert model parameters and computations to int8. Table \ref{tab:Others model} compares our sm2 model with baseline and state-of-the-art methods, highlighting its superior performance.

\begin{table}[h]
\centering

\caption{Performance under Various Parameter Configurations}
\label{tab:proposed-models}
\centering

\begin{tabular}{>{\centering\arraybackslash}p{0.06\linewidth}>{\centering\arraybackslash}p{0.09\linewidth}>{\centering\arraybackslash}p{0.09\linewidth}>{\centering\arraybackslash}p{0.09\linewidth}>{\centering\arraybackslash}p{0.09\linewidth}>{\centering\arraybackslash}p{0.1\linewidth}>{\centering\arraybackslash}p{0.1\linewidth}}
\hline
 \textbf{\textbf{System}}&   \textbf{Param/K} &   \textbf{MACs/M} &    \textbf{Size/KB}&\textbf{ACC/\%}&\textbf{Size(QAT)}&\textbf{ACC(QAT)}\\ \hline
 \textbf{sm1}& 13.76 & 3.88&    55.03
&56.54
&   13.76 &   53.44\\
 \textbf{sm2}& 30.69 & 8.27&    177.79
&61.42
&   30.69
&   58.25\\
 \textbf{sm3}& 54.98 & 14.34 &    219.93
&63.03&   54.98&   60.29\\
 \textbf{sm4}& 126.12  & 23.03  &    504.47 &65.26&   126.12&   62.11\\
\hline
\end{tabular}

\end{table}

\begin{table}[h]
\centering
\caption{Comparing Model Performance}
\label{tab:Others model}
\begin{tabular}{lccc}
\hline
  \textbf{Model}& \textbf{Param/K} &\textbf{MACs/M} &\textbf{ACC} \\
\hline
  \textbf{Baseline} \cite{Martínmorato2022lowcomplexity}& 46.51&29.23&42.9\\
 
  \textbf{CP-ResNet} \cite{Cai2023}& 63.5& 19.5&57.0\\
   \textbf{BSConv-CNN} \cite{Tan}& 73.39& 13.18&55.6\\
  \textbf{Proposed Model (sm2)}& \textbf{30.69} &\textbf{8.27}&\textbf{58.25}\\
\hline
\end{tabular}
\end{table}

\begin{table*}[h]
\centering
\caption{Performance Comparison of Fused and Averaged Teacher Models: the comparison experiment is based on the sm4 model}
\label{tab:teacher}
{
\small

\begin{tabular}{l|ccc|cccccc|c}
\hline
 \textbf{Methods}&\textbf{A}&\textbf{B}&\textbf{C}&\textbf{S1} & \textbf{S2}& \textbf{S3}& \textbf{S4}& \textbf{S5}&\textbf{S6} &\textbf{ACC}\\
\hline
 Teacher(w/o)&69.77&64.27& 64.90& 58.24& 56.67& 60.38& 52.43& 58.30& 49.36&59.36\\
 Averaged&71.38&62.60&66.50& 65.20& 59.53& 65.78& 62.05& 61.53& 56.71&63.48\\

 \textbf{Fused}& \textbf{74.00}& \textbf{65.29}& \textbf{68.64}& \textbf{66.44}& \textbf{61.31}& \textbf{67.02}& \textbf{62.71}& \textbf{62.90}& \textbf{59.04}&\textbf{65.26}\\
 \hline
\end{tabular}}
\end{table*}

\begin{table*}[h]
\centering

\caption{Performance Comparison: the comparison experiment is based on the sm4 model.\textbf{ACC} represents the average validation set accuracy over the last ten rounds, while  \textbf{A}, \textbf{B}, and \textbf{C}  indicate accuracy on real devices. S1-S3 and S4-S6 denote accuracy on simulated devices observed and unseen during training, respectively. }
\label{tab:conbined}
\resizebox{\textwidth}{!}{
\begin{tabular}{l|ccc|cccccc|c}
\hline
 \textbf{Model}&\textbf{A}&\textbf{B}&\textbf{C}&\textbf{S1} & \textbf{S2}& \textbf{S3}& \textbf{S4}& \textbf{S5}&\textbf{S6} &\textbf{ACC}\\
\hline
 DS-FlexiNet+FMS+DIR&71.34&64.30& 66.43& 64.26& 60.85& 65.72& 62.26& 62.69& 56.88&63.86\\
  DS-FlexiNet+BN+FMS+DIR& 70.57& 63.76& 66.97& 64.55& 59.91& 65.65& 60.68& 61.63& 56.70&63.38\\
  DS-FlexiNet+BN+FMS+ADIR& 73.70& \textbf{66.19}& 68.54& 64.75& 62.06& 66.23& 61.76& 62.42& 56.45&64.68\\
  DS-FlexiNet+RN+FMS+DIR& 72.55& 64.12& 68.18& 65.36& \textbf{62.48}& 66.89& 61.41& \textbf{62.92}& 57.35&64.58\\
 \textbf{DS-FlexiNet+RN+FMS+ADIR}&\textbf{74.00}&65.29&\textbf{68.64}& \textbf{66.44}& 61.31& \textbf{67.02}& \textbf{62.71}& 62.90& \textbf{59.04}&\textbf{65.26}\\
\hline
\end{tabular}}
\end{table*}

\subsection{Performance Optimization and Resource Utilization}

The DS-FlexiNet model excels on the TAU Urban Acoustic Scenes 2022 Mobile Dataset, outperforming traditional models in classifying audio from unknown devices such as S4, S5, and S6.  Additionally, QAT optimizes the model by reducing storage and computational costs while enhancing performance with limited data, making it ideal for mobile and embedded systems.

\subsection{Dimensional Analysis of Residual Normalization }

Residual Normalization (RN) significantly improves acoustic scene classification accuracy and robustness across various recording environments and devices. By normalizing along the channel dimension, RN reduces misclassification between similar scenes, such as parks and street traffic, and mitigates channel-specific variations more effectively than frequency normalization. RN achieves up to 86\% accuracy in classifying scenes like bus, park, and street traffic, as shown in Fig.~\ref{fig:RN}(a) and Fig.~\ref{fig:RN}(b).

Unlike BN, which can lead to blurring of subtle differences, RN preserves channel-specific information, enhances scene distinction, and reduces instability from batch data variations, leading to superior performance and stability in ASC.

\subsection{Comparing Fused and Averaged Teacher Models }

Table \ref{tab:teacher} demonstrates the benefits of using fused logits to guide the student model. Compared to the simpler averaging method with multiple teacher models, the fusion strategy dynamically adjusts weights and biases to leverage the strengths of each teacher model. 

Through fusing teacher models, this strategy enhances the student model's understanding and generalization capabilities. It enables effective adaptation to different devices and resource limitations, leading to improved performance in ASC across diverse scenarios.

\subsection{Enhancing Model Robustness and Generalization}

Table \ref{tab:conbined} shows significant accuracy improvements with audio data from various devices, especially in the test set. RN enhances generalization by reducing the impact of recording environments and background noise, particularly in unseen scenarios. ADIR boosts adaptability by preserving key characteristics and minimizing distortion. Together, RN and ADIR greatly improve the model's robustness and generalization across diverse environments and unknown devices.

 \section{Conclusion }

In summary, our method achieves notable accuracy improvements across various devices, particularly in challenging test datasets. RN, applied along the channel dimension, and data augmentation, including ADIR and FMS, effectively tackle cross-device generalization by mitigating noise, preserving key characteristics, and reducing distortion. The integration of depthwise separable convolutions, QAT, and KD ensures DS-FlexiNet delivers high performance with reduced computational demands, making it ideal for resource-constrained deployments. These innovations address both cross-device variability and efficiency in low-resource conditions.

\section*{Acknowledgment}

We are  grateful to the organizers of the Detection and Classification of Acoustic Scene and Events challenge for providing us with access to a large amount of open-source audio data.

\end{document}